\documentclass[english,aps,pra,reprint,noshowpacs,nofootinbib,superscriptaddress]{revtex4-1}
\pdfoutput=1
\usepackage[T1]{fontenc}	% should generally be included for better accented-word behavior
\usepackage[latin9]{inputenc}	% should generally be included for better accent behavior
\usepackage{geometry}		% for controlling page margins
\usepackage{graphicx}
\usepackage[above,below]{placeins}	% allows use of \FloatBar­rier command to force section barriers
\usepackage{times}
\usepackage{hyperref}
\hypersetup{colorlinks=true,urlcolor=blue,citecolor=blue,linkcolor=blue}
\begin{document}

\title{How computation can facilitate sensemaking about physics: A case study}
\author{Odd Petter Sand}
\author{Tor Ole B. Odden}
\author{Christine Lindstr\o m}
\affiliation{Centre for Computing in Science Education (CCSE), University of Oslo, N-0316 Oslo, Norway}
\author{Marcos D. Caballero}
\affiliation{Centre for Computing in Science Education (CCSE), University of Oslo, N-0316 Oslo, Norway}
\affiliation{Department of Physics and Astronomy and Create for STEM Institute, Michigan State University, East Lansing, MI 48823}

%\keywords{Sensemaking, computation}

\begin{abstract}
We present a case study featuring a first-year bio-science university
student using computation to solve a radioactive decay problem and
interpret the results. In a semi-structured cognitive interview, we use
this case to examine the process of sensemaking in a computational science
context. We observe the student entering the sensemaking process by
inspecting and comparing computational outputs. She then makes several
attempts to resolve the perceived inconsistency, foregrounding knowledge
from different domains. The key to making sense of the model for this
student proves to be thinking about how to implement a better model
computationally. This demonstrates that integrating computation in
physics activities may provide students with opportunities to engage in
sensemaking and critical thinking. We finally discuss some implications
for instruction.
\end{abstract}

\maketitle

\section{Introduction}

It is a well-known problem that students can progress through
introductory physics courses, sometimes with good grades, and still lack
understanding of the underlying principles, relations, and concepts. A
common scenario is that students employ ``plug and chug'' strategies to
manipulate mathematical formulae without engaging with the underlying
physical principles. With this in mind, getting students to engage in
sensemaking is crucial for achieving learning goals in critical thinking
and understanding the physics itself \cite{ref1}.

Computation is important for students of physics to learn because it
reflects current practices in the field, teaches important skills for
research and other careers, and allows students to solve a greater
number of more realistic problems \cite{ref2}. Consequently,
research-based efforts to sensibly integrate computation into the
physics curriculum are well underway \cite{ref3, ref4, ref5}. Therefore, we
want to study to what extent computation provides a potential for
students engaging in sensemaking, and under which conditions that
potential may be fully realised.

We present evidence for sensemaking in the case of Sophia, a bio-science
student who is interviewed while solving a physics problem on
radioactive decay. Sophia uses both computational and non-computational
arguments to make sense of the model. The process of modifying her program
and comparing the outputs turns out to facilitate Sophia's sensemaking.
We justify this claim by presenting evidence for how computation was helpful
in the sensemaking process. Finally, we discuss implications for teaching and
future research.

\section{Analytical Framework}

The analytical framework for this study is founded on the following
definition of sensemaking from \cite{ref6}, pp. 5-6: ``A dynamic process of
building or revising an explanation in order to {[}\ldots{}{]} resolve a
gap or inconsistency in one's understanding.'' While there have been
numerous other attempts to define what sensemaking is, we chose this one
because it unifies several aspects of sensemaking that others have
highlighted: sensemaking as an epistemological frame, a cognitive
process, and a discourse practice, all of which are relevant to this
project.

The process of sensemaking involves (a) realising that there is a gap or
contradiction in one's knowledge, (b) iteratively proposing ideas and
attempting to connect them to existing knowledge or other ideas, and (c)
evaluating that these ideas are consistent and do not lead to additional
contradictions \cite{ref6}. In this paper, we will use this definition to
study how computational activities may provide opportunities for
sensemaking in interdisciplinary science problems.

\section{Methods}

The case comes from a pilot study conducted with first-year bio-science
students at a large research-intensive university in Norway. These
students learned computation integrated with biology in the previous
semester and were following a physics course in the semester when this
study took place. The physics course had not yet covered radioactive
decay by the time we interviewed the students. We targeted students with
a wide range of self-reported programming expertise who were also
comfortable thinking aloud.

Subsequently, we performed a series of semi-structured cognitive
interviews in Norwegian where students worked on the task alone. The
interviews borrowed heavily from think-aloud protocols, but students
could ask for help with syntax should they need it, provided they were
able to articulate what they wanted the code we gave them to do.

Follow-up questions on students' reasoning were asked by the interviewer
on various occasions, interspersed throughout the think-aloud segments.
This tends to change the students' thought processes, often improving
the results. Protocols obtained in this way tend to be more valid than
the ones were students recall their reasoning after the fact, however
\cite{ref7}.

We gave the interviewees a toy model starting off with 1000 radioactive
nuclei and told them that 10\% of the remaining nuclei would decay every
month. The students first calculated the remaining number of nuclei for
the first two months (where the answers were still integers) by hand.
The next step for the students was to reproduce these answers by writing
a Python program in Jupyter Notebook. This is the familiar programming
environment they used throughout the previous semester. Finally, they
were asked to extend the calculations to 60 and 100 months and (if time
allowed) plot the results.

This task was specifically designed to allow students to discover a
perceived trade-off between accuracy and realism that would require
sensemaking to resolve. After a while, you need several decimal points
to mathematically describe 10\% of what remains, yet when counting
nuclei, in general one expects the numbers to be integers. While the toy
model we provided may be approximately correct for a large number of
nuclei, at lower amounts one would have to interpret the output as an
average across many identically prepared experiments for the numbers to
make sense.

All the students interviewed (N=5) at some point considered rounding the
answers to the closest integer to avoid working with fractions of
nuclei, although some did this only in response to follow-up questions
from the interviewer. Every student also expressed some 
concern about the mathematical accuracy of their results when rounding
the numbers in this way. Two of the interviewees made some progress
toward resolving this contradiction by interpreting the un-rounded
numbers as an average, one of which was Sophia.

The typical length of an interview was about one hour. All interviews
were recorded on audio and video, both of the student and the computer
screen. Subsequently, the transcripts were translated from Norwegian
into English. We analysed the transcripts using the definition in
\cite{ref6} and looked for the following: The student (a) realising she
cannot fully explain the physical phenomenon she is modelling or aspects
of the model itself, (b) proposing explanations and trying to connect
them to scientific or everyday knowledge and (c) evaluating these
explanations to ensure consistency.

We then looked at what the student was doing with computation inside and
outside of these sensemaking episodes, and asked the following
questions: What happens in this computational context when the student
engages in sensemaking? Is the computational aspect of the task a help
or hindrance to this process?

The case we present illustrates how sensemaking may happen in a
computational context. While not the most typical case for this group of
students, Sophia's interview was chosen for analysis because her
sensemaking was rather explicit in the transcript. Additionally, she
ended up using language that was clearly computational to make a
profound argument about how to model the physical phenomenon and
interpret the results.

\section{Computational Sensemaking Case}

``Sophia'' (pseudonym) is a Norwegian student in her mid-20s, a few
years older than most students taking first-year university courses. She
describes her experience with programming as one of a fair degree of
mastery in most cases. Compared to the average student in the
programming course for bio-science students, she comes across as more
confident and relaxed than most when working with computer code. During
the interview, she rarely asked for confirmation that she was on the
right track, and she did not hesitate long before trying something out.

We begin our analysis at the point where Sophia has set up her program
to calculate the number of remaining nuclei for the first three months:
1000, 900.0 and 810.0, respectively.

\begin{quote}
\textbf{Sophia {[}14:35{]}} \emph{There. Now it's right. {[}But{]} now I might
want to round these} {[}indicates 900.0 and 810.0{]} \emph{to get\ldots{} well,
just whole numbers.}
\end{quote}

She implements this rounding to the closest integer when displaying the
output from the program, but not in the actual calculations, and checks
that it works.

\begin{quote}
\textbf{Interviewer {[}15:05{]}} \emph{Could you tell me a little more about
why you would want to round them?}

\textbf{Sophia} \emph{Because these are atoms, and you sort of can't have
half\ldots{} or I don't know\ldots{} it seems a little unnecessary to
include, like, 810.0 atoms, in a way.}
\end{quote}

We interpret \emph{``you sort of can't have half\ldots{}''} as that you
cannot have a fraction of a nucleus and still call it a nucleus of that
particular element, which is a point Sophia returns to later on.

At this point, we have reached the start of the sensemaking process. It
is divided into three separate segments that correspond to the three
ideas Sophia proposes to make physical sense of the numbers given to her
by her program.

\subsection{Sensemaking segment I}

Sophia moves on to the next part of the task, modifying her program to
repeat calculations all the way up to 60 months. She inspects the output
and indicates the last ten months in the sequence, with 3, 3, 3, 2, 2,
2, 2, 2, 1, and 1 nucleus, respectively.

\begin{quote}
\textbf{Sophia {[}16:30{]}} \emph{This looks a little strange\ldots{} Because
here there are no decimals. So\ldots{} here I'd include the decimals
because, like\ldots{} you can't take 10 percent of\ldots{} or, I get
that you get, like, the same number several months in a row.}
{[}indicates the earlier sequence 6, 6, 5, 5{]} \emph{Because 10 percent of 6
is still above 5, like. I'm going to include the decimals.}
\end{quote}

While cutting the decimals for large numbers seems fine to her, Sophia
realises that for smaller numbers there is something she needs to find
an explanation for. Why does the number of
nuclei remain constant for several time steps and then changes more
than 10\% rather abruptly? In terms of our sensemaking framework, the
sensemaking process thus starts in reaction to the computational output
when she realises that something is \emph{``a little strange''}.

Using computation also allows her to include the decimals and test this
change, which she immediately does. Yet, in terms of our framework, the
idea that Sophia proposes here is first and foremost mathematical. She
talks about numbers in a sequence, decimals and percentages, but this
discussion stands on its own removed from the physics and computational
contexts it occurred in.

\subsection{Sensemaking segment II}

After resolving some bugs (one syntax error and a few logical errors),
Sophia sees the un-rounded numbers for all 60 months. After verifying
that they seem to be the correct numbers mathematically, she is told
that she is free to move on to the next task. Despite this suggestion,
she decides she would rather continue making sense of the model.

\begin{quote}
\textbf{Sophia {[}20:18{]}} \emph{Umm, yes. Right now, I'm thinking -- I just
have to say it, because right now I am a little unsure about\ldots{}
because there are now so many decimals and\ldots{}} {[}indicates the
final months with 2.21\ldots{}, 1.99\ldots{} and 1.79\ldots{} nuclei{]}
\emph{because one atom can't\ldots{} you can't take 10 percent of one atom,
like. So, this becomes sort of random whether, in a way\ldots{} whether
it splits or, like, if it loses one atom to radioactivity or not. So,
I'm really not entirely happy with these numbers. But I can move on to
the next one, I guess.}
\end{quote}

We interpret this as Sophia revisiting her earlier statement: Can you
have a fraction of a nucleus? The outstanding feature of this segment is
the critique of her previous choice, which according to our framework is
indicative of sensemaking going on.

Initially, Sophia seems hesitant to exit the
sensemaking process prematurely, and
she may be experiencing some friction between the sensemaking and how
she frames the interview situation. The initial ``\emph{I just have to
say it}'' at 20:18 seems to indicate that at that point she was about to
engage in an activity she considered not wholly appropriate for the way
she was framing the activity at the time \cite{ref8}.

One should also note that in contrast to the previous sensemaking
attempt, this one foregrounds the ideas from physics (atoms,
radioactivity) with a nod to the mathematics embedded in them
(percentages, randomness).

\subsection{Sensemaking segment III}

At this point the interviewer intervenes and invites Sophia to discuss a
little more why she is not happy with the numbers, in effect sustaining
the sensemaking frame. Initially this invitation is met with minor
resistance, possibly because it was suggested she move on
in segment II. Sophia states that she does not want to spend so much
time and energy thinking about an open-ended task which is not clear
about what it wants from her, so she is \emph{``choosing the easy way
out''}. After being asked what she would do if she were a scientist and
this was an important result to her, Sophia resumes the sensemaking
process:

\begin{quote}
\textbf{Sophia {[}23:20{]}} \emph{So, already after the third month here, then
I would have taken, like,} {[}indicates month 4 with 656.1 nuclei{]} \emph{here
it reads point 1 -- then I might have put in a} for \emph{loop with} choice? \emph{I
think it is} {[}random.choice()\footnote{\url{https://docs.python.org/3/library/random.html\#random.choice}}{]} \emph{
you use. Whether or not, like, that one\ldots{} like, whether the
decimal, whether that is a whole atom that goes away or not. So, in a
way it becomes a sort of choice\ldots{} thing. Such that when you run it
as a model for the first time, then maybe\ldots{} yes. Then maybe
all\ldots{} eh, the radioactive atoms are spent after, like, 56
months\ldots{} and then the next time they are spent after 60 months.
And the time after that maybe after 70 months. Eh, and then I
would\ldots{} yes, then I would have made a program or maybe a}
def-\emph{function and then run that many times and look at, percentage-wise,
then, how probable is it that, eh, all the atoms\ldots{} yeah, are gone
after 50 months or after 70 months. So, I'd rather make that kind of
model, because\ldots{} eh, you kind of can't make this} {[}indicates the
output{]} \emph{completely accurate... But at the same time, when I think
about it, it is\ldots{} the probability of when that is going to happen
is a little present in these numbers, too.}
\end{quote}

At this point Sophia is using \emph{computation} as a tool for
sensemaking, something that was not explicitly evident in her earlier
attempts. The mathematics and physics are still present in the
background. Referring back to our framework once again, Sophia did
mention randomness in segment II, yet this is the first time she
proposes to interpret each un-rounded number as an average. But
critiquing that idea leads to the question: an average of what? Of
different \emph{simulations.} \emph{``I would have made a program}
{[}\ldots{}{]} \emph{and then run that many times and look at,
percentage-wise, then, how probable is it that} {[}\ldots{}{]} \emph{all
the atoms} {[}\ldots{}{]} \emph{are gone after 50 months or after 70
months.''} We claim that this point, firmly embedded in the
computational nature of the task, is key for Sophia's bridging the gap
in her understanding she has been wrestling with.

As opposed to the simplified difference equation she was working with
originally, the approach suggested here incorporates randomness: two
sets of 1000 nuclei would not necessarily decay in identical ways. This
realisation does not mean she has a complete idea of how to implement it
computationally, but sensemaking is about how you get there.

In summary, we have identified three sensemaking segments, in which
Sophia foregrounds knowledge from the following domains:

\begin{itemize}
\item Segment I: Mathematics
\item Segment II: Physics
\item Segment III: Computation
\end{itemize}

These segments together clearly demonstrate the sensemaking
process: Sophia (a) realises that rounding the numbers hides
information. It seems inaccurate that the number of nuclei appears
unchanged for several time steps and then abruptly changes significantly
more than 10\%. But \emph{not} rounding the numbers leads to working
with fractions of a nucleus, which conflicts with her intuition about
how the world works, as established prior to segment I. In each segment
Sophia (b) iterates by proposing ideas and (c) critiquing these to make
sure they are consistent in themselves and with other ideas.

The sensemaking process ends with the resolution of changing the
interpretation of the numbers in the toy model. Instead of the actual
number of nuclei in one experiment they represent an average across an
ensemble of computational simulations. At this point, we interpret
Sophia's statements to mean that she regards both integers and decimal
numbers as valid outputs from her program. She has also attained a rough
idea of how to implement the simulations in question.

\section{Discussion and Conclusions}

In this paper, we have shown that computation helped Sophia in two ways.
First, she was able to modify her program back and forth between
rounding and no rounding with relative ease. In the first two
sensemaking segments, inspecting and comparing the outputs of these
approaches provides an entry point into the sensemaking process: \emph{``This
looks a little strange\ldots{}''}

Second, we argue that the key to Sophia's interpretation of her output
as an average is to think computationally about the problem, which is
what happens in segment III. When discussing how to implement a more
realistic model computationally, she realises that her current results
can be interpreted as an average of several such simulations:
\emph{``The probability of when that is going to happen is a little
present in these numbers, too.''}

Without claiming that this case is common or representative
for this group of students, we argue
that this case study provides an existence proof that computation can
provide fertile ground for student engaging in sensemaking.
Specifically, working computationally allowed Sophia to (a) realise a
gap in her understanding, (b) implement ideas and (c) test and critique
the results for consistency. We observed that in this context, the idea
that drew most heavily on computational knowledge proved the most
fruitful in the sensemaking process.

To determine under which circumstances this potential for sensemaking
can be fulfilled, further research is needed. In the other four
interviews, we did note other examples of students beginning to engage
in sensemaking in response to the output of their programs. What is
special about Sophia's case was the way her computational resources
helped her make sense of the apparent contradiction between the physics
(realism) and mathematics (accuracy) in the model. She also initially ignored the
interviewer's suggestion to move on at the start of segment II. It
remains to investigate how this would play out in a classroom setting,
where there is no interviewer to help sustain the sensemaking process
like at the start of segment III. Video observations is one possible way
to probe this.

Future studies could also identify the thresholds for entering and
successfully resolving a sensemaking process, respectively,
using computation. This would have profound implications for how instructors
integrate computation in science classes, for instance when designing
tasks that go beyond procedural use of computer programming as a tool.
If critical thinking is important to us, we should attempt to
realise the full sensemaking potential in computational activities. It
is then necessary to ensure that our students have sufficiently strong
computational foundations to engage in these sensemaking tasks.

\acknowledgments{This study was funded by the Norwegian Agency for Quality Assurance in
Education (NOKUT), which supports the Centre for Computing in Science
Education.}

% For short, simple bibliographies, manually formatting works:
% Remember that you'll need to run pdfLaTeX twice to get the references to show (first
% pass will insert ?? in their places).

% For a longer bibliography, delete the thebibliography block above, then comment in
% these two lines to use a .bib file with BibTeX.
%\bibliographystyle{apsrev}  	% supercedes the longbibliography option, so leave commented out if you want to display article titles
%\bibliography{mybibfile}  	% don't include the .bib suffix

\end{document}